\input lanlmac
\input epsf.tex
\input mssymb.tex
\overfullrule=0pt

\newcount\figno
\figno=0
\def\fig#1#2#3{
\par\begingroup\parindent=0pt\leftskip=1cm\rightskip=1cm\parindent=0pt
\baselineskip=11pt
\global\advance\figno by 1
\midinsert
\epsfxsize=#3
\centerline{\epsfbox{#2}}
\vskip 12pt
{\bf Fig.\ \the\figno:} #1\par
\endinsert\endgroup\par
}
\def\figlabel#1{\xdef#1{\the\figno}%
\writedef{#1\leftbracket \the\figno}%
}
\def\omit#1{}

\def\pre#1{{\tt
#1}}
\def\qed{\nobreak\hfill\vbox{\hrule height.4pt%
\hbox{\vrule width.4pt height3pt \kern3pt\vrule width.4pt}\hrule height.4pt}\medskip\goodbreak}
\lref\RS{A.V. Razumov and Yu.G. Stroganov, 
{\sl Combinatorial nature
of ground state vector of $O(1)$ loop model},
{\it Theor. Math. Phys.} 
{\bf 138} (2004) 333-337; {\it Teor. Mat. Fiz.} 138 (2004) 395-400, \pre{math.CO/0104216}.}
\lref\BdGN{M.T. Batchelor, J. de Gier and B. Nienhuis,
{\sl The quantum symmetric XXZ chain at $\Delta=-1/2$, alternating sign matrices and 
plane partitions},
{\it J. Phys.} A34 (2001) L265--L270,
\pre{cond-mat/0101385}.}
\lref\FR{I.B.~Frenkel and N.~Reshetikhin, {\sl Quantum affine Algebras and Holonomic 
Difference Equations},
{\it Commun. Math. Phys.} 146 (1992), 1--60.}
\lref\DFZJ{P.~Di Francesco and P.~Zinn-Justin, {\sl Around the Razumov--Stroganov conjecture:
proof of a multi-parameter sum rule}, {\it E. J. Combi.} 12 (1) (2005), R6,
\pre{math-ph/0410061}.}
\lref\Pas{V.~Pasquier, {\sl Quantum incompressibility and Razumov Stroganov type conjectures},
\pre{cond-mat/0506075}.}
\lref\ZJDF{P.~Di Francesco and P.~Zinn-Justin, {\sl Quantum Knizhnik--Zamolodchikov equation,
generalized Razumov--Stroganov sum rules and extended Joseph polynomials}, 
{\it J. Phys. A } {\bf 38}
(2006)  L815-L822, \pre{math-ph/0508059}.}
\lref\DF{P.~Di Francesco, {\sl A refined Razumov--Stroganov conjecture},
J. Stat. Mech. P08009 (2004), \pre{cond-mat/0407477}.}
\lref\STEM{J. Stembridge, {\sl Nonintersecting paths, Pfaffians and plane partitions},
Advances in Math. {\bf 83} (1990) 96-131.}
\lref\Bre{D. Bressoud, {\sl Proofs and Confirmations: The Story of
the Alternating Sign Matrix Conjecture}, Cambridge Univ. Pr., 1999.}
\lref\Ku{G. Kuperberg, {\sl Symmetry classes of alternating-sign
matrices under one roof}, 
{\it Ann. of Math.} 156 (3) (2002), 835--866,
\pre{math.CO/0008184}.}
%
%
\lref\LGV{B. Lindstr\"om, {\it On the vector representations of
induced matroids}, Bull. London Math. Soc. {\bf 5} (1973)
85-90\semi
I. M. Gessel and X. Viennot, {\it Binomial determinants, paths and
hook formulae}, Adv. Math. { \bf 58} (1985) 300-321. }
\lref\KM{A.~Knutson and E.~Miller, {\sl Gr\"obner geometry of Schubert polynomials},
{\it Annals of Mathematics} (2003), \pre{math.AG/0110058}.}

\Title{T-06/084}
{\vbox{
\centerline{Totally Symmetric Self-Complementary Plane Partitions}
\medskip
\centerline{and Quantum Knizhnik-Zamolodchikov equation:}
\medskip
\centerline{a conjecture}
}}
\bigskip\bigskip
\centerline{P.~Di~Francesco} 
\medskip
\centerline{\it  Service de Physique Th\'eorique de Saclay,}
\centerline{\it CEA/DSM/SPhT, URA 2306 du CNRS,}
\centerline{\it F-91191 Gif sur Yvette Cedex, France}
\bigskip
\vskip0.5cm
\noindent
We present a new conjecture relating the minimal polynomial solution
of the level-one $U_q(\frak{sl}(2))$ quantum Knizhnik-Zamolodchikov equation for generic
values of $q$ in the link pattern basis and some $q$-enumeration of Totally Symmetric 
Self-Complementary Plane Partitions.

\bigskip

\Date{07/2006}
%
%
\newsec{Introduction}

Statistical physics have always nurtured a particular relationship with 
enumerative combinatorics. A remarkable manifestation
of this fact is the recent conjecture by Razumov and Stroganov (RS) \RS,
relating the solution of the O(1) dense loop model on a semi-infinite cylinder
of square lattice with perimeter $2n$ to the refined enumeration of $n\times n$ 
Alternating Sign Matrices (ASM) in the (bijectively equivalent) form of configurations 
of the Fully-Packed Loop (FPL) model on an $n\times n$ square grid.

An actual proof \DFZJ\ of a weaker {\it sum rule} version of the RS conjecture \BdGN\ was actually
constructed by making full use of the integrability of the O(1) model. The latter allowed to
completely determine the vector of (renormalized) probabilities for the configurations 
of an inhomogeneous version of the model on the cylinder to connect boundary points according 
to given connectivity patterns, and the sum rule followed. An alternative proof was found in
\Pas, in the course of studying polynomial representations of the affine Temperley-Lieb algebra.
The latter provide a deformation of the original model, involving a parameter $q$.
This was recast in \ZJDF\ as the problem of finding non-trivial {\it minimal polynomial} 
solutions of the quantum Knikhnik-Zamolodchikov 
(qKZ) equation for the level one quantum enveloping algebra $U_q(\frak{sl}(2))$ \FR, 
the particular ``RS" value of $q$ being $q=-e^{i\pi/3}$.
It was also noted in \ZJDF\ that another particular value of $q$, $q=-1$, actually contains
lots of combinatorial wonders. Indeed, when taking a suitable $q\to -1$ 
limit of the homogeneous version of the above probabilities, one ends up with non-negative integers
which were shown to count the degrees of the components of the variety $M^2=0$, where $M$ is a
complex strictly upper triangular matrix, thus providing another RS-type theorem, 
in relation with algebraic geometry.

In the homogeneous case but with $q$ generic, one gets a remarkable sum rule for the 
abovementioned probabilities, in the form of polynomials of the variable 
\eqn\deftau{\tau=-q-q^{-1}}
with apparently only {\it non-negative integer} coefficients.
The aim of this note is to provide a conjectural answer for what these coefficients
actually count.

In the long story of the ASM conjecture (see Bressoud's book \Bre\
for a complete saga and references), many puzzles remain open, one of which is the 
still elusive relation 
between ASM of size $n\times n$ and the Totally Symmetric 
Self-Complementary Plane Partitions (TSSCPP) of size $2n$, or alternatively the
rhombus tilings of a regular hexagon of triangular lattice, with side of length $2n$,
with all symmetries of the hexagon and also the condition that is be identical to its
complement when interpreted as the view in perspective of a piling up of unit cubes 
within a cube of size $2n$. Despite many efforts, no natural bijection has yet been
found between ASM and TSSCPP. It is however rather simple to enumerate TSSCPP, via a 
bijection to a set of Non-Intersecting Lattice Paths (NILP), which in turn are 
counted by determinant formulas.

In this note, we present the following conjecture:
\item{}
{\bf The weighted enumeration of TSSCPP of size $2n$, in the form of 
lattice paths with two 
types of steps, and with a weight $\tau=-q-q^{-1}$ per step of the first type produces the 
generic $q$ sum rule for the above minimal polynomial solution of the level-one 
$U_q(\frak{sl}(2))$ qKZ equation.}
\par

If true, this conjecture provides a new relation between TSSCPP and ASM, although not 
yet one-to-one, but rather many-to-many. It also suggests that the RS conjecture may
be extended to one that allows to interpret the abovementioned probabilities in the 
homogeneous case as weighted counting functions for restricted classes of TSSCPP, as
each of these quantities also appears to be a polynomial of $\tau$ with non-negative
integer coefficients. This ultimate refinement would pave the way to a natural TSSCPP-ASM
bijection, yet to be found.

The paper is organized as follows. In Section 2, we review briefly the enumeration of
TSSCPP by reinterpreting them as NILP. The latter is refined by introducing a weight $\tau$
per vertical step, leading to polynomials $P_{2n}(\tau)$, also expressed 
via a simple Pfaffian formula. In Section 3, we review briefly the minimal polynomial
solution of the level-one $U_q(\frak{sl}(2))$ qKZ equation in the link pattern basis, 
and concentrate on the sum of components in the homogeneous case, leading to polynomials 
$\Pi_{2n}(\tau)$. This leads to the main conjecture of the paper, namely that
$P_{2n}(\tau)=\Pi_{2n}(\tau)$ (Sect. 4.1) which we check at $\tau=2$ and then at
generic $\tau$ by numerically
solving the qKZ equation by means of a new algorithm, based on the decomposition of the
solution as a sum of products of monomials of a particular type (Sect. 4.2).
The main consequences of the conjecture as well as additional remarks are gathered
in the concluding Section 5.

\newsec{TSSCPP}

\fig{A typical TSSCPP of size $2n=12$. The area delimited by a black broken line
is a fundamental domain for all symmetries of the partition ($1/12$th of the hexagon).
This domain is expressed as a configuration of NILP, by following the sequences
of rhombic tiles ${\epsfxsize=0.4cm\vcenter{\hbox{\epsfbox{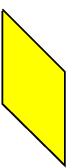}}}}$ 
and ${\epsfxsize=0.4cm\vcenter{\hbox{\epsfbox{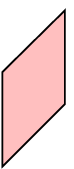}}}}$.}{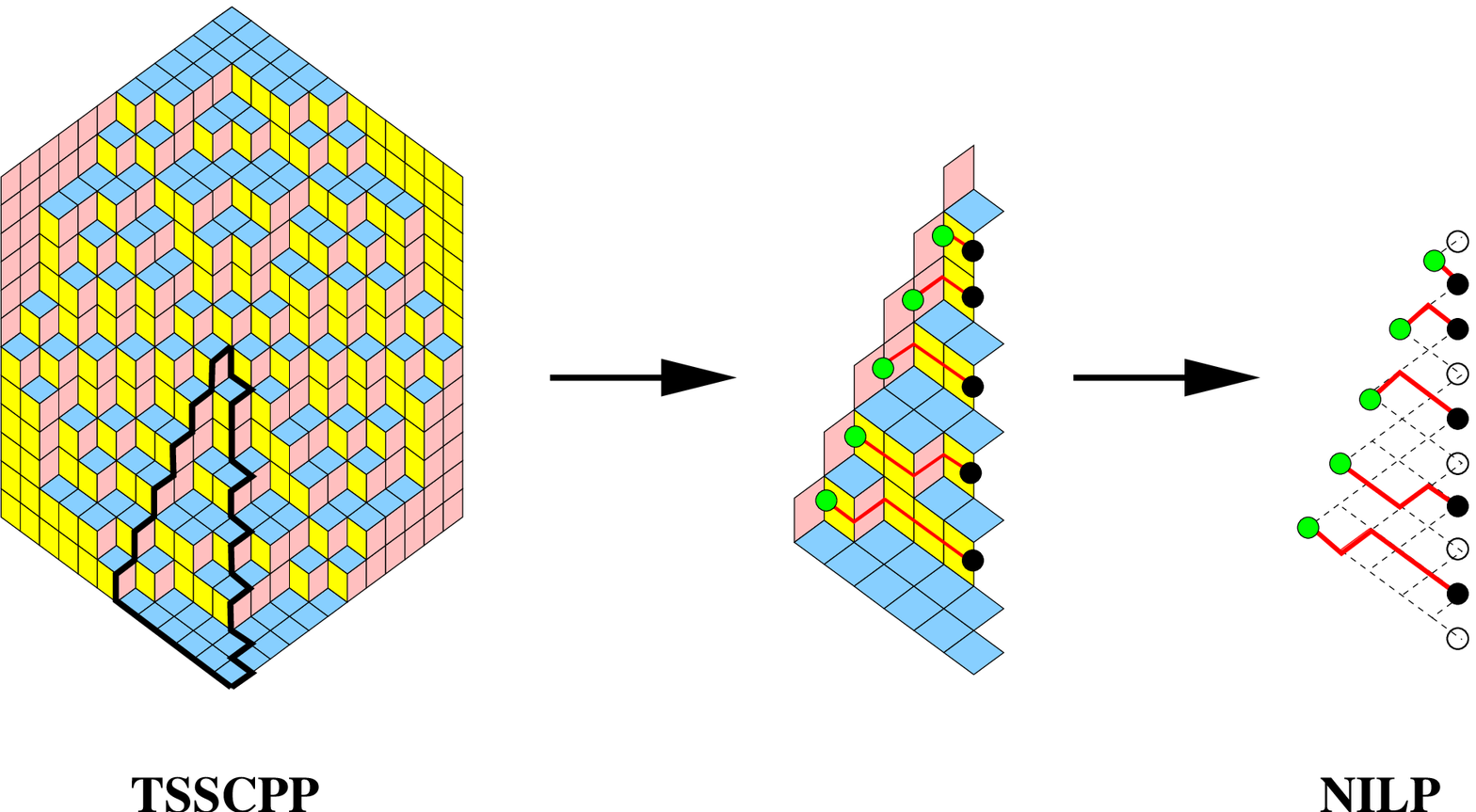}{13.cm}
\figlabel\tsscppnilp

The counting of TSSCPP may be best performed by using their formulation as NILP.
To make a long story short, we simply give a pictorial representation of the
TSSCPP-NILP correspondence in Fig.\tsscppnilp. Starting from a TSSCPP in the 
from of a totally symmetric self-complementary rhombus tiling of a regular hexagon 
of triangular lattice with edge of length $2n$, we concentrate on a fundamental domain
made of $1/12$th of the hexagon, and note that its tiling configuration is entirely
determined by the configurations of $n-1$ sequences of rhombi of two
of the three types used (see middle picture in Fig.\tsscppnilp). 
The latter form NILP, which are represented on the right of Fig.\tsscppnilp.

\fig{A sample NILP for $n=6$. The paths make only up $(0,1)$ and right-diagonal
$(1,1)$ steps on the underlying square lattice, and start from the points
$(i,-i)$ (green dots) and end up on the x axis (red dots), at points
$(r_i,0)$, for $i=1,2,\ldots 5$.}{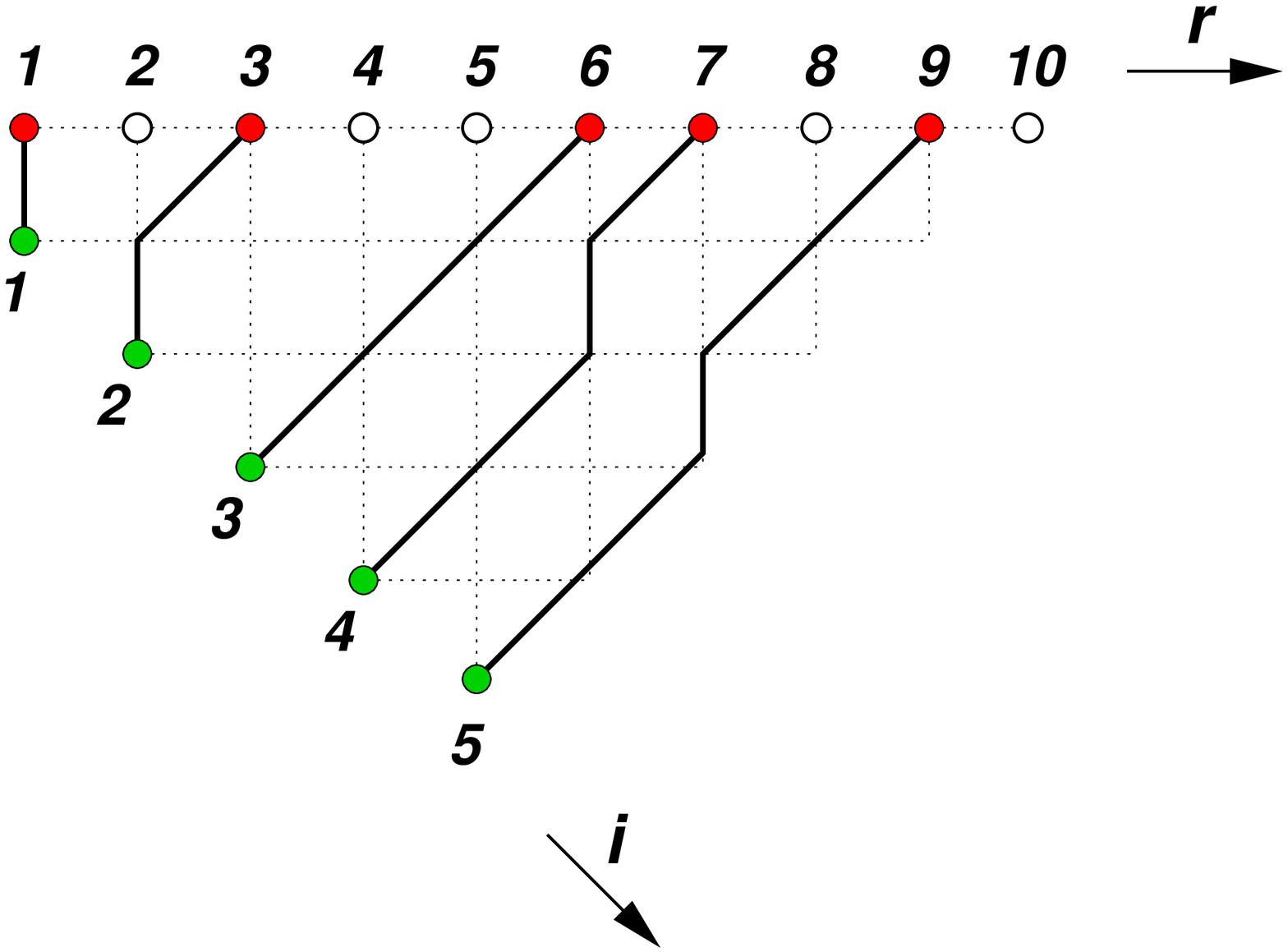}{6cm}
\figlabel\nilpsamp

Upon trivial deformations of the underlying triangular lattice and harmless rotations
and reflexions, the problem boils down to that of counting of NILP on the square lattice, 
starting from the points $(i,-i)$, $i=1,2,\ldots,n-1$ and ending at positive integer points
on the x axis, of the form $(r_i,0)$, $i=1,2,\ldots,n-1$, and making ``vertical"
steps $(0,1)$ or ``diagonal" ones $(1,1)$ only (see Fig.\nilpsamp\ for illustration). 
It is clear that the $i$-th path,
starting from $(i,-i)$ must end at $(r_i,0)$, as the paths do not intersect each other.
Moreover, we must have $r_i\leq 2i$ for all $i$, and the sequence $r_i$ is strictly
increasing.

The counting of paths with prescribed ends $r_i$, $i=1,2,\ldots,n-1$ is readily
performed by use of the Lindstr\"om-Gessel-Viennot (LGV) formula \LGV, expressing
the total number of NILP configurations as a determinant:
\eqn\lgvnilp{ P_{2n}(r_1,r_2,\ldots,r_{n-1})=\det_{1\leq i,j\leq n-1} {\cal P}(i,r_j)}
where ${\cal P}(i,r)$ denotes the number of paths from $(i,-i)$ to $(r,0)$. This latter
number is nothing but
\eqn\notbut{ {\cal P}(i,r)={i\choose r-i} }
as we simply have to choose $r-i$ diagonal steps among a total of $i$. Note that 
the number of vertical steps is therefore $i-(r-i)=2i-r$. Hence, if we wish to 
obtain the weighted enumeration of TSSCPP with prescribed endpoints $r_i$
and with a weight $\tau$ per vertical
step, we simply have to multiply ${\cal P}(i,r)$ by $\tau^{2i-r}$ in \lgvnilp.
This yields the weighted enumeration of TSSCPP with prescribed endpoints:
\eqn\weienum{ P_{2n}(r_1,r_2,\ldots,r_{n-1}\vert \tau)=\det_{1\leq i,j\leq n-1}
\left( {i\choose r-i} \tau^{2i-r} \right)}
and the total weighted sum of TSSCPP of size $2n$ reads then
\eqn\totwei{ P_{2n}(\tau)=\sum_{1\leq r_1<r_2<\cdots <r_{n-1}\atop
r_i\leq 2i } P_{2n}(r_1,r_2,\ldots,r_{n-1}\vert \tau) }
It is a polynomial in $\tau$ of degree $n(n-1)/2$, the total number of steps 
of each configuration.

For completeness we list below the first few values of this polynomial.
\eqn\firfew{\eqalign{ P_{2}(\tau)&=1\cr
P_4(\tau)&=1+\tau\cr
P_6(\tau)&=1+3\tau+2\tau^2+\tau^3\cr
P_8(\tau)&=1+6\tau +11\tau^2+12\tau^3+8\tau^4+3\tau^5+\tau^6\cr
P_{10}(\tau)&=1 + 10\tau + 35\tau^2 + 70\tau^3 + 98\tau^4 + 
91\tau^5 + 69\tau^6 + 35\tau^7 + 15\tau^8\cr
&+4\tau^9 + \tau^{10}\cr
P_{12}(\tau)&=1 + 15\tau + 85\tau^2 + 275\tau^3 + 628\tau^4 + 
1037\tau^5 + 1346\tau^6 + 1379\tau^7 + 1144\tau^8\cr
&+783\tau^9 + 435\tau^{10} + 204\tau^{11} + 74\tau^{12} + 
24\tau^{13} + 5\tau^{14} + \tau^{15}\cr}}

\fig{The case of size $2n=6$. We have represented the TSSCPP with fixed numbers
of vertical steps, ranging from $0$ to $3$ (indicated in the top row), leading to
the polynomial $P_6(\tau)=1+3\tau+2\tau^2+\tau^3$.}{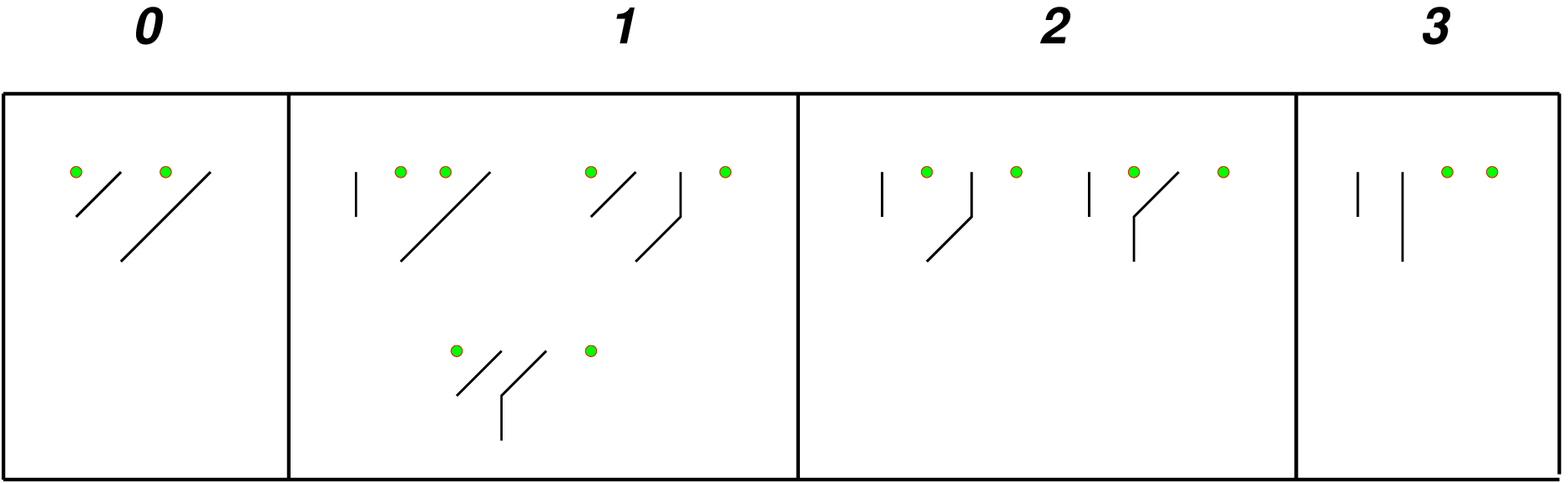}{12.cm}
\figlabel\conjsix

\fig{The case of size $2n=8$. We have represented the TSSCPP with fixed numbers
of vertical steps, ranging from $0$ to $6$ (indicated in the top row), leading to
the polynomial 
$P_8(\tau)=1+6 \tau +11 \tau^2+12\tau^3+8\tau^4+3\tau^5+\tau^6$.}{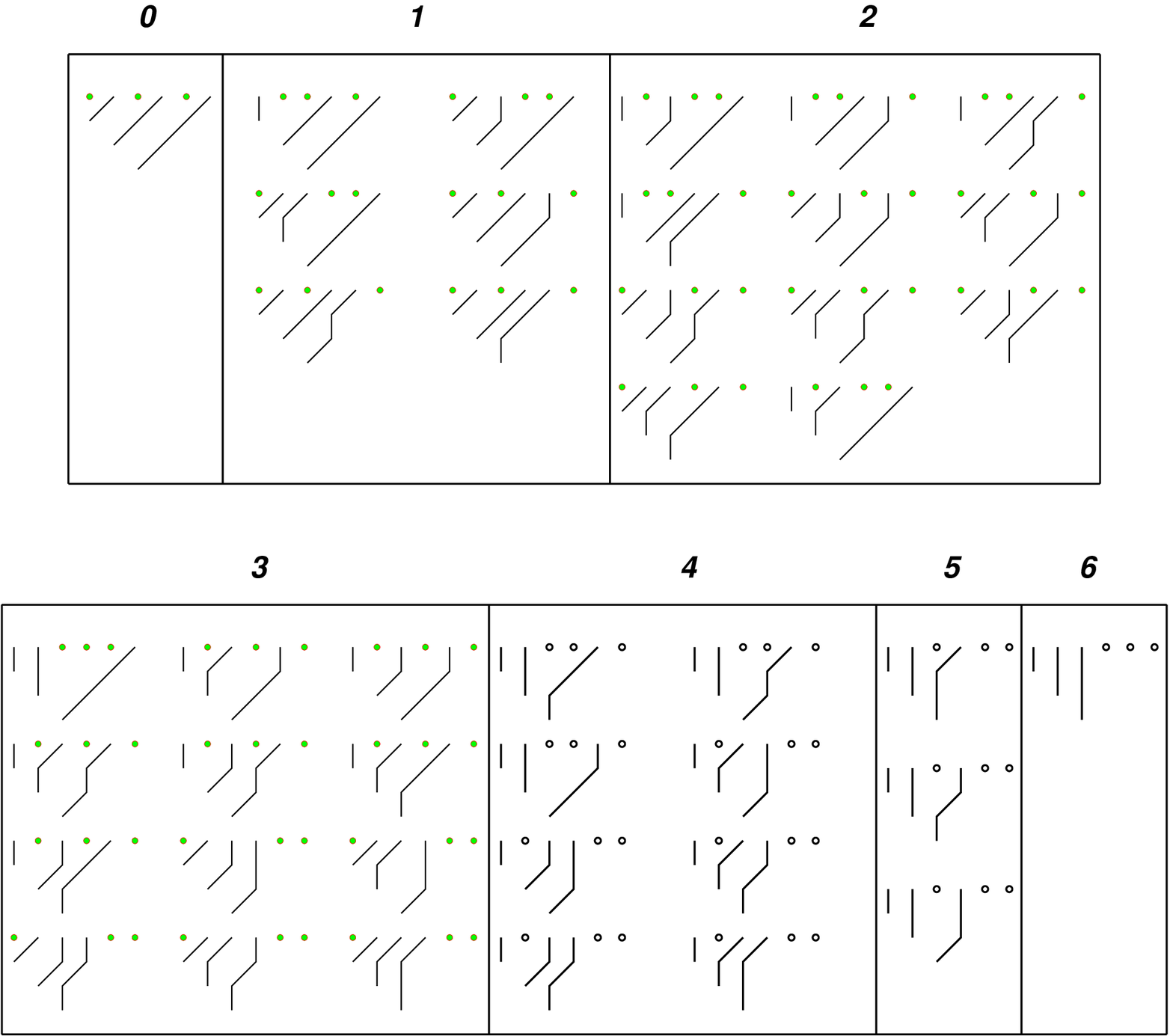}{13.cm}
\figlabel\conjeight

For illustration, we have represented in Figs. \conjsix\ and \conjeight\ 
the TSSCPP of size $2n=6$ and $8$ respectively,
arranged according to their numbers of vertical steps.

At $\tau=1$, $P_{2n}(1)$ reproduce the celebrated TSSCPP numbers usually denoted by 
$N_{10}(2n,2n,2n)$, also equal to the 
total numbers of ASM of size $n\times n$:
$N_{10}(2n,2n,2n)=P_{2n}(1)=A_n=\prod_{0\leq j\leq n-1} {(3j+1)!\over (j+n)!}$,
with values $1,2,7,42,429,\ldots$ for $n=1,2,3,4,5,\ldots$

At $\tau=-1$, we get an interesting sequence $1,0,-1,0,9,0,-646,\ldots$ 
for $n=1,2,3,4,5,6,7\ldots$,
which we have been able to 
(conjecturally) relate to
the so-called refined ASM numbers $A_{n,k}$ (see \Bre\ and references therein), 
counting ASM of size $n\times n$ with the unique $1$ 
in the first row at position $k$, with values
\eqn\refnumasm{ A_{n,k}={n+k-2\choose k-1} {(2n-k-1)!\over (n-k)!} 
\prod_{j=0}^{n-2}{(3j+1)!\over (n+j)!} }
The observed relation reads:
\eqn\relarefasm{ P_{2n}(-1)=\sum_{k=1}^n (-1)^{{n+1\over 2}-k} A_{n,k}}
Note that this vanishes when $n$ is even, and that moreover when $n$ is odd,
the result is a perfect square up to a sign, which we have identified as:
\eqn\lutfin{ P_{4n+2}(-1)=(-1)^n A_V(2n+1)^2}
where $A_V(2n+1)$ is the number of Vertically Symmetric ASM (VSASM),
with value \Ku:
\eqn\valvsasm{ A_V(2n+1)=\prod_{j=0}^{n-1} {(6j+4)!(2j+2)!\over (4j+4)!(4j+2)!}}
We have been able to check the relation \lutfin\ for $n$ up to $6$, i.e. for TSSCPP of size
up to $26$.

It has been known for quite some time that the TSSCPP are also enumerated via a
Pfaffian expression \STEM, simply expressing in physical terms the 
fermionic character
of non-intersecting lattice paths. We have found an easy way to generalize the Pfaffian
expression to the present case. Let $\epsilon_n=(1-(-1)^n)/2$. Then introducing
the quantity
\eqn\qty{ H(i,j\vert \tau)=\sum_{i\leq r<s \leq 2j} \tau^{2i+2j-r-s}
\left({i\choose r-i}{j\choose s-j}-{i\choose s-i}{j\choose r-j}\right) }
the polynomial $P_{2n}(\tau)$ is nothing but the Pfaffian:
\eqn\pfarel{ P_{2n}(\tau) 
={\rm Pf}\, \left(H(i,j\vert \tau)\right)_{\epsilon_{n}\leq i<j \leq n-1}}
In particular, the above identification at $\tau=-1$ implies a new Pfaffian formula
for the square of the number of VSASM, yet to be proved:
\eqn\newpfa{ A_V(2n+1)^2=(-1)^n {\rm Pf}\left(
\sum_{i\leq r<s \leq 2j} (-1)^{r+s}
\left({i\choose r-i}{j\choose s-j}-{i\choose s-i}{j\choose r-j}\right)
 \right)_{1\leq i<j\leq 2n} }

\newsec{Minimal polynomial solution of the qKZ equation}

The inhomogeneous O(1) dense loop model on a semi-infinite cylinder of
square lattice of perimeter $2n$ was investigated in Ref.\DFZJ. In particular, 
the probability ${\cal P}_\pi$ for a loop configuration 
to connect the $2n$ boundary points by pairs via non-intersecting links according
to the link pattern $\pi$ was computed. Here a link pattern is simply a chord
diagram on a disk with $2n$ regularly spaced boundary points, numbered 
$1$ to $2n$ counterclockwise, connected by pairs via $n$ non-intersecting chords.
There are $c_n={(2n)!\over n!(n+1)!}$ such link patterns.
These correspond to the effective connections between the points on the boundary 
of the semi-infinite cylinder, via bulk configurations of non-intersecting links.
The inhomogeneous character of the model is simply via Boltzmann weights that
depend on the vertical position of the face configurations, via complex parameters
$z_1,z_2,\ldots, z_{2n}$.
In \DFZJ, the probabilities ${\cal P}_\pi$ were renormalized up to a global constant
by a common overall polynomial factor in such a way that they all be polynomials 
of the $z$'s with no common factors. We denote by 
$\Psi_\pi\equiv \Psi_\pi(z_1,z_2,\ldots z_{2n})$ these renormalized probabilities.
The quantities $\Psi_\pi$ were shown to obey the following system of equations
\eqn\sispsi{\eqalign{
\tau_{i,i+1} \Psi(z_1,\ldots,z_{2n})&= {\check R}_{i,i+1}(z_i,z_{i+1}) 
\, \Psi(z_1,\ldots,z_{2n}),\quad
i=1,2,\ldots 2n-1\cr
\sigma \,\Psi(z_2,\ldots,z_{2n},z_1)&=\Psi(z_1,\ldots,z_{2n})\cr}}
where we have used the following notations: $t_{i,i+1}$ for the elementary transposition
$z_i\leftrightarrow z_{i+1}$; ${\check R}_{i,i+1}$ for the R-matrix operator
acting linearly on the link pattern basis as
\eqn\rmatact{ {\check R}_{i,i+1}= {q^{-1}z_i-qz_{i+1}\over q^{-1}z_{i+1}-qz_{i}}Id+
{z_i-z_{i+1} \over q^{-1}z_{i+1}-qz_{i}}e_i }
with $q=-e^{i\pi/3}$, and 
where $e_i$ is the $i$-th generator of the Temperley-Lieb algebra acting on link patterns
as follows: $e_i$ creates a link between points $i$ and $i+1$ while it glues the links
connected to $i$ and $i+1$ respectively (any loop created in the process may be safely
erased, as the loop weight is $\tau=-q-q^{-1}=1$); finally $\sigma$ acts on link patterns
as a counterclockwise rotation by one unit.

The system \sispsi\ may be alternatively written in components as
\eqn\compopsi{\eqalign{
(q^{-1}z_{i+1}-q z_i)\partial_i \Psi_\pi &=\sum_{\pi'\neq \pi\atop
e_i \pi' =\pi } \Psi_{\pi'}, \quad \forall \pi\in \ {\rm Im}(e_i),  \ 1\leq i\leq 2n-1 \cr
\Psi_\pi(z_2,\ldots,z_{2n},z_1)&=\Psi_{\sigma(\pi)}(z_1,\ldots,z_{2n})\cr}}
where we have introduced the divided difference operator
\eqn\dividif{ \partial_i f(z_i,z_{i+1})= {f(z_{i+1},z_i)-f(z_i,z_{i+1})\over
z_i -z_{i+1}} }
which decreases by $1$ the degree of the polynomial $f$. These equations were then
solved by noticing first that all components $\Psi_\pi$ may be obtained in a triangular
way from the fundamental component $\Psi_{\pi_0}$ corresponding to the link pattern
$\pi_0$ that connects points $i\leftrightarrow 2n+1-i$, $i=1,2,\ldots,n$, and
then that the latter reads:
\eqn\fundapsi{ \Psi_{\pi_0}=\prod_{1\leq i<j\leq n} (qz_i -q^{-1}z_j) \, 
\prod_{n+1\leq i<j\leq 2n} (qz_i -q^{-1}z_j) }

In \DFZJ, an expression for the sum rule $\sum_\pi \Psi_\pi$ was derived based on
symmetry properties and recursion relations inherited from eq.\compopsi. It was then
realized first in the language of affine Temperley-Lieb algebra \Pas, and then in that
of the qKZ equation \ZJDF, that a natural extension of this holds for generic $q$ as well,
at the expense of giving up the loop gas picture. Indeed, the system
\sispsi\ is nothing but a particular case of qKZ equation for the value $q=-e^{i\pi/3}$.
The generic $q$ equation reads
\eqn\genepsi{\eqalign{
\tau_{i,i+1} \Psi(z_1,\ldots,z_{2n})&= {\check R}_{i,i+1}(z_i,z_{i+1}) 
\, \Psi(z_1,\ldots,z_{2n}),\quad
i=1,2,\ldots 2n-1\cr
\sigma \,\Psi(z_2,\ldots,z_{2n},q^6z_1)&=q^{3(n-1)}\, \Psi(z_1,\ldots,z_{2n})\cr}}
with ${\check R}_{i,i+1}$ and $\sigma$ as before, but for $q$ generic 
(note that the $e_i$ are now generators of the Temperley-Lieb algebra 
with weight $\tau$ per loop, as opposed to $1$ before).
Remarkably the generic $q$ minimal polynomial solution to these equations is
essentially the same, namely that every component is obtained triangularly in terms
of $\Psi_{\pi_0}$ by successive actions with divided difference operators
and multiplications by monomials, and that the expression for $\Psi_{\pi_0}$
remains the same as in \fundapsi, but now with $q$ generic.
For illustration, in the case $n=3$, there are $5$ link patterns with $2n=6$
boundary points (which we conveniently represent along a line
rather than on a circle, as the cyclic invariance is now lost, and we find that
\eqn\solqkzsix{\eqalign{
&\Psi_{\epsfxsize=1.8cm\vcenter{\hbox{\epsfbox{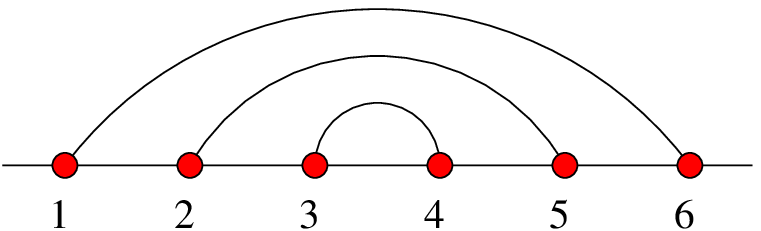}}}}
=a_{1,2} a_{1,3}a_{2,3}a_{4,5}a_{4,6}a_{5,6}\cr
&\Psi_{\epsfxsize=1.8cm\vcenter{\hbox{\epsfbox{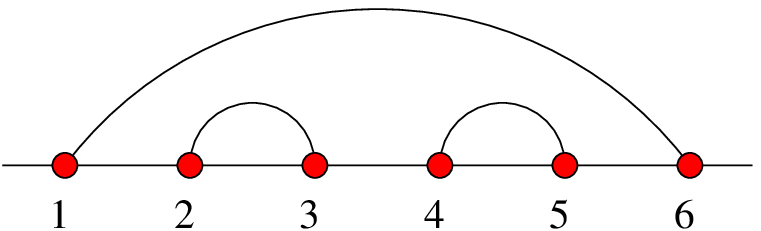}}}}
=a_{1,2}a_{3,4}a_{5,6}(a_{1,3}a_{4,6}b_{5,2}+a_{2,4}a_{3,5}b_{6,1})\cr
&\Psi_{\epsfxsize=1.8cm\vcenter{\hbox{\epsfbox{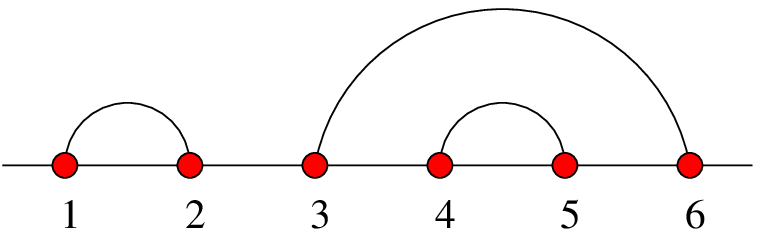}}}}
=a_{2,3}a_{2,4}a_{3,4}a_{5,6}b_{5,1}b_{6,1}\cr
&\Psi_{\epsfxsize=1.8cm\vcenter{\hbox{\epsfbox{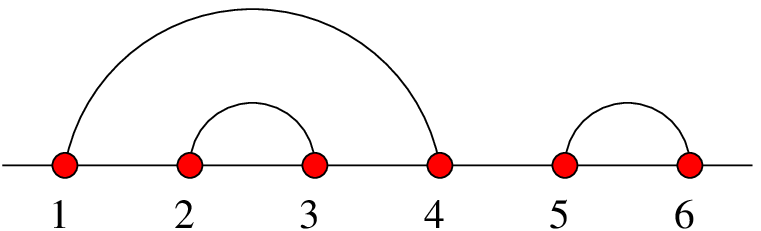}}}}
=a_{1,2}a_{3,4}a_{3,5}a_{4,5}b_{6,1}b_{6,2}\cr
&\Psi_{\epsfxsize=1.8cm\vcenter{\hbox{\epsfbox{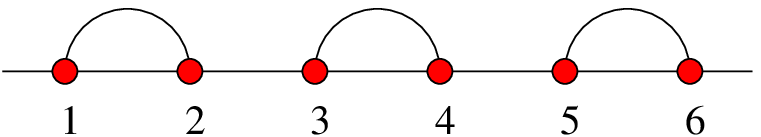}}}}
=a_{2,3}a_{4,5}b_{6,1}(a_{2,4}b_{5,1}b_{6,3}+a_{1,2}a_{3,5}a_{4,6})\cr}}
Here we have introduced a convenient notation: 
\eqn\nota{a_{i,j}=qz_i-q^{-1}z_j, \qquad {\rm and} \qquad
b_{i,j}=q^{-2}z_i-q^2 z_j}
Note that these two definitions coincide at the RS 
point $q=-e^{i\pi/3}$. 

We may then consider the renormalized probabilities $\Psi_{\pi}$ in the homogeneous
limit where all the $z_i$'s are taken to $1$, while $q$ remains fixed and generic.
Actually, the correct choice of normalization in this case is really to
consider $\varphi_\pi=\Psi_{\pi}/\Psi_{\pi_0}$, so that $\varphi_{\pi_0}=1$.
It turns out that all quantities $\varphi_\pi$ are polynomials of $\tau=-q-q^{-1}$
with non-negative integer coefficients. Consequently, the sum rule
\eqn\sumrulqkz{ \Pi_{2n}(\tau)\equiv
\sum_\pi {\Psi_\pi(1,1,\ldots,1)\over \Psi_{\pi_0}(1,1,\ldots,1)}}
also produces polynomials of $\tau$ with non-negative integer coefficients.

Returning to the example $n=3$ of eq.\solqkzsix, and noting that $b_{i,j}=\tau a_{i,j}$ 
in the homogeneous limit, we find, upon dividing the entries above by 
$\Psi_{\pi_0}=(q-q^{-1})^6$, that 
\eqn\homopsi{\eqalign{
&\varphi_{\epsfxsize=1.8cm\vcenter{\hbox{\epsfbox{arch0.eps}}}}=1\cr
&\varphi_{\epsfxsize=1.8cm\vcenter{\hbox{\epsfbox{arch1.eps}}}}=2 \tau\cr
&\varphi_{\epsfxsize=1.8cm\vcenter{\hbox{\epsfbox{arch2.eps}}}}=\tau^2\cr
&\varphi_{\epsfxsize=1.8cm\vcenter{\hbox{\epsfbox{arch3.eps}}}}=\tau^2\cr
&\varphi_{\epsfxsize=1.8cm\vcenter{\hbox{\epsfbox{arch4.eps}}}}=\tau+\tau^3\cr}}
henceforth $\Pi_6(\tau)=1+3\tau+2\tau^2+\tau^3$. 
Note that the powers of $\tau$ are simply equal to the numbers of $b$ factors in 
each product of $a$'s and $b$'s forming the quantities $\Psi_\pi$.
Note also the coincidence of $\Pi_6(\tau)$ with $P_6(\tau)$ of eq.\firfew.

\newsec{The conjecture}

\subsec{The conjecture}

We now have all the elements to formulate the following

\noindent{\bf Conjecture:}

The polynomial $P_{2n}(\tau)$, equal to the generating function for TSSCPP of size $2n$, 
with a weight $\tau$ per vertical step (in NILP form),
and the polynomial $\Pi_{2n}(\tau)$, equal to the homogeneous sum rule 
for the solution of the level-one $U_q(\frak{sl}(2))$ qKZ equation at generic $q$, 
do coincide for all $n$, with $\tau=-q-q^{-1}$.

\subsec{Checks and useful algorithms}

A first non-trivial check of the above conjecture regards the values of the polynomials 
at $\tau=2$.
This corresponds to $q=-1$, where earlier findings \ZJDF\ have shown that the 
quantities $\varphi_\pi$ may be interpreted as the degrees of the components of
the variety $M^2=0$, $M$ an upper triangular complex matrix of size $2n\times 2n$.
In particular, the total degree $d_{2n}=\Pi_{2n}(2)$ of this variety, 
should equal $P_{2n}(2)$.
Computing this total degree numerically, we have found a nice formula for $d_{2n}$ matching
the results for sizes up to $12$. It is again a sum of determinants, which actually also
counts yet another set of non-intersecting lattice paths. It reads:
\eqn\fortotdeg{ d_{2n}=\sum_{1\leq r_1<\ldots <r_{n-1} \atop r_i \leq 2i}
\det_{1\leq i,j\leq n-1} {2i \choose r_j} }
and enumerates NILP that start at points $(2i,0)$, end up at points $(r_i,r_i)$, and
take only vertical steps $(0,1)$ and horizontal steps $(-1,0)$,
with $i=1,2,\ldots,n-1$.

To show that $d_{2n}=P_{2n}(2)$,
let us now prove that the rectangular $n-1\times 2n-2$ matrices $B$ and $A$
with entries $B_{i,r}={2i \choose r}$ and $A_{i,r}={i\choose r-i} 2^{2i-r}$ actually share
the same minors of size $n-1\times n-1$. To prove this, it is sufficient to find
a square matrix $Q$ of determinant $1$, such that $B=QA$. It turns out
that the lower triangular matrix with entries $Q_{k,i}={k \choose i}$ does the job.
Indeed, let us compute the following generating function for the entries of $QA$:
\eqn\genfunqa{\eqalign{ a_k(x)&\equiv\sum_{i,r\geq 1}
x^r {k\choose i} {i\choose r-i}2^{2i-r}\cr
&=\sum_{i\geq 1}(2x)^i {k\choose i} \sum_{m\geq 0} {i\choose m}
\left({x\over 2}\right)^{m} \cr
&=\sum_{i\geq 1}(2x+x^2)^i {k\choose i}=(1+x)^{2k}-1\cr}}
henceforth, picking the coefficient of $x^r$ for $r\geq 1$, we end up with $B=QA$.

We have checked the general conjecture for arbitrary $\tau$ up to $n=6$. The
computation of $P_{2n}(\tau)$ is straightforward, thanks to the formula \pfarel.
On the other hand, that of the sum rule $\Pi_{2n}(\tau)$ is more subtle, as it involves
constructing the solution of the qKZ equation explicitly, and only in the end
taking the homogeneous limit. To do this, we have found a powerful algorithm, that might
eventually help prove the RS conjecture. It is based on the following simple remark on
the master equation 
\eqn\mast{(q^{-1}z_{i+1}-q z_i)\partial_i \Psi_\pi=
\sum_{\pi'\neq \pi\atop e_i \pi' =\pi } \Psi_{\pi'},\quad
i=1,2,\ldots 2n-1}
when $\pi$ is in the image of $e_i$,
which allows to compute all $\Psi_\pi$'s in a triangular way in terms of $\Psi_{\pi_0}$.

\fig{The bijection between link patterns and Dyck paths in size $2n=6$.
The corresponding inclusion order on
link patterns reads: $\pi_0<\pi_1<\pi_2,\pi_3<\pi_4$.}{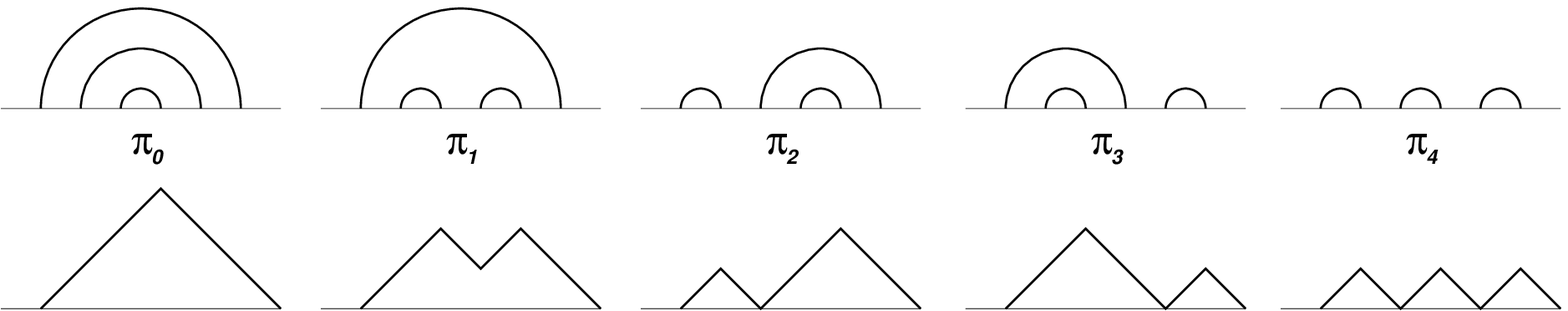}{13.cm}
\figlabel\linkpatodyck

This goes as follows. We first represent the link patterns as a set of
$n$ non-intersecting semi-circles connecting $2n$ regularly spaces points along a line,
labeled from $1$ to $2n$, and then use a standard bijection 
to Dyck paths (see Fig.\linkpatodyck\ for
the case $2n=6$), namely
paths of length $2n$ in the plane starting from the origin and taking only diagonal steps 
$(1,1)$ or $(1,-1)$, ending at $(2n,0)$, and visiting only points $(x,y)$ with $y\geq 0$.
The bijection is obtained as follows: visiting the points of the link pattern from left to
right, we form a Dyck path by taking a $i$-th step $(1,1)$ if a semi-circle
originates from the point $i$ or $(1,-1)$ if a semi-circle terminates at point $i$. 
We may now define an order on link patterns $\pi<\pi'$ iff the Dyck path of $\pi$ 
{\it contains} that of $\pi'$, namely for each $i$-th visited
point $(x_i,y_i)$ and $(x_i',y_i')$ of the respective paths, 
the difference $(y_i-y_i')\geq 0$, and at least one of these quantities
is non-zero. Now given a link pattern $\pi$ in the image
of $e_i$, its antecedents $\pi'\neq \pi$ under $e_i$ are all such that $\pi'< \pi$,
except for one, say $\pi^*$, such that $\pi^*>\pi$. In the Dyck path language, $\pi^*$
is identical to $\pi$ except that its steps $i$ and $i+1$ are interchanged. Therefore,
triangularly w.r.t. inclusion of Dyck paths, we may determine recursively
\eqn\deterpsi{ \Psi_{\pi^*}=(q^{-1}z_{i+1}-q z_i)\partial_i \Psi_\pi 
-\sum_{\pi'\neq \pi,\pi^*\atop e_i \pi' =\pi } \Psi_{\pi'} }

The main difficulty here is to actually evaluate the l.h.s. and to put it in a suitable
form. We have found that each quantity $\Psi_\pi$ may be written as a {\it sum of products}
of $a_{i,j}$ and $b_{k,l}$'s of eq.\nota, with $i<j$ and $k>l$, without any additional 
numerical factor (see eq.\solqkzsix\ for an example at $2n=6$). The problem is that
the expressions as sums are not unique, but their homogeneous values are, even at
generic $q$. So the algorithm simply expresses the $\Psi_\pi$, order by order in
the computation, as such sums of products. More precisely, we start
from 
\eqn\fundab{ \Psi_{\pi_0}=\prod_{1\leq i<j\leq n}a_{i,j} \, \prod_{n+1\leq i<j\leq 2n}a_{i,j}}
and write eq.\mast\ for $i=n$ (which is the only place where the link pattern $\pi_0$
has a chord linking points $i$ and $i+1$, hence is the image
under $e_i$ of some other link pattern).
Acting with $\partial_i$ on any symmetric polynomial of $z_i,z_{i+1}$ gives zero by definition,
hence $\partial_n$ has a non-trivial action only on the part of $\Psi_{\pi_0}$
that is not symmetric in $z_n,z_{n+1}$, namely on $\prod_{i=1}^{n-1} a_{i,n}
\prod_{j=n+2}^{2n}a_{n+1,j}$. To proceed, we now need two lemmas on the action
of $\partial_i$ on monomials.

\noindent{\bf Lemma 1:}

The divided difference operator obeys a modified Leibnitz rule:
\eqn\modilei{ \partial_i (f g)= \partial_i(f) g+t_{i,i+1}(f)\partial_i(g) }
as a straightforward consequence of the definition \dividif.

\noindent{\bf Lemma 2:}
The operator $T_i\equiv (q^{-1}z_{i+1}-q z_i)\partial_i$ acts as follows on products of two
monomials:
\eqn\actontwomon{\eqalign{
T_i (a_{n,i}\, a_{i+1,p})&=a_{i,i+1}\, b_{p,n},\quad {\rm for}\ n<i<i+1<p \cr
T_i (a_{p,i}\, b_{i+1,n})&=a_{i,i+1}\, a_{n,p},\quad {\rm for}\ n<p<i<i+1 \cr
T_i (b_{n,i}\, a_{i+1,p})&=a_{i,i+1}\, a_{n,p},\quad {\rm for}\ i<i+1<n<p \cr
}}

We may now explain the algorithm: starting from a $\Psi_\pi$ already
expressed as a sum of products of $a$'s and $b$'s, apply the operator $T_i$ on
each such product. First group the terms of the product that are non-symmetric in $z_i,z_{i+1}$
into pairs of the form $a a$ or $a b$ like in the l.h.s. of eq.\actontwomon (it turns out
that this can always be done, possibly in several ways).
Then use the modified Leibnitz rule \modilei\ to express the action of 
$T_i$ as a sum of terms in which 
$T_i$ only acts on one pair. Use \actontwomon\ to express this action. We are left
with a sum of products of $a$'s and $b$'s. The only difficulty is to make sure
indices in the pairs are ordered as in \actontwomon, but then the action is straightforward.
There are unfortunately many choices of pairs, some of which may lead to dead ends,
where no use of \actontwomon\ may be allowed, hence the algorithm must explore many
possibilities before converging. 

A time-saving remark however is that we have not made use of the quasi-cyclicity relation
\eqn\quasicyc{q^{-3(n-1)}\Psi_\pi(z_2,\ldots,z_{2n},q^6z_1)=\Psi_{\sigma(\pi)}(z_1,\ldots,z_{2n})}
which is actually granted once we pick $\Psi_{\pi_0}$ as in eq.\fundapsi.
It is easy to see that this relation may be applied to any product of $a$ and $b$ monomials,
in which the index $2n$ appears exactly $n-1$ times (a fact easily checked for
the products of $a$'s and $b$'s we generate), and that under the action of $\sigma$
we have
$a_{i,j}\to a_{i+1,j+1}$ unless
$j=2n$, in which case $a_{i,2n}\to b_{i+1,1}$, and $b_{i,j}\to b_{i+1,j+1}$,
unless $j=2n$, in which case $b_{i,2n}\to a_{1,i+1}$. So the quasicyclic invariance condition
\quasicyc\ allows to generate all the rotated versions $\Psi_{\sigma^r(\pi)}$ as sums of products 
of $a$'s and $b$'s out of an expression for $\Psi_\pi$ as a sum of products of $a$'s and $b$'s.
Combining this with the main algorithm, we have been able to generate solutions of qKZ
up to size $2n=12$, and found perfect agreement with the conjecture of Sect. 4.1.

\newsec{Consequences and conclusion}

In this note we have presented a new conjecture relating TSSCPP in the form of NILP 
and ASM or generalized sum rules for the qKZ equation.
The algorithm developed in Sect. 4.2 above, might help find a bijection
between ASM and TSSCPP. Indeed, comparing the generic $q$ situation with that of the RS point, 
where there is no more distinction
between $a$'s and $b$'s, and assuming the RS conjecture to be true,
there are exactly as many products of $a$'s and $b$'s forming $\Psi_\pi$
at generic $q$ as there are FPL configurations with the link pattern $\pi$. 
So these expressions (unfortunately not unique at this point, but one may hope 
for the existence of a canonical decomposition) of $\Psi_\pi$ as
sums of products of $a$'s and $b$'s suggest to attach each such product to each FPL
configuration with link pattern $\pi$ on one hand, and to some TSSCPP with as many vertical
steps as there are $b$'s in the product, on the other hand. Crossing this information
with that of refined ASM and TSSCPP of Ref.\DF, this might eventually lead to 
a bijection between ASM and TSSCPP, and perhaps to a proof of the RS conjecture.

Another interesting consequence of the conjecture regards the point $q=-1$, namely $\tau=2$.
As mentioned in Sect. 4.2, the quantity $\Pi_{2n}(2)=d_{2n}$ counts the total
degree of the variety $M^2=0$ for upper triangular complex matrices $M$.
$P_{2n}(2)$, when expressed as a sum over NILP yields via the above conjecture
a formula for the total degree $d_{2n}$
as a sum of powers of $2$ (each $a$ contributes $1$ and each $b$ contributes $2$
in the homogeneous case $z_i=1$ followed by the $q\to -1$ limit of $\varphi_\pi$),
each term in the sum corresponding to a TSSCPP. 
The TSSCPP therefore play the role of ``pipe dreams" \KM\ for the degree
counting, and suggest that the variety $M^2=0$ may be decomposed into complete intersections of
linear hyperplanes and quadratic (degree 2) varieties, and that this decomposition involves
exactly as many terms (components) as the number of ASM or TSSCPP. Such an algebro-geometric
interpretation of ASM or TSSCPP numbers would be quite nice.

Another point of interest is $q=e^{i\pi/3}$, i.e. $\tau=-1$, where our conjectured relation 
between $P_{4n+2}(-1)$ and the square of the number of VSASM
of size $(2n+1)\times (2n+1)$, also yields via the main conjecture of this paper a relation
between the numbers $F_{n}(\pi)$ of FPL configurations of size 
$n\times n$ with link pattern $\pi$, 
and the number $A_V(n)$. Indeed, the quantities $\Psi_\pi$ enjoy some parity property
that all products of $a$'s and $b$'s in their decomposition have numbers of $b$ of the same parity.
Hence in the homogeneous case, the corresponding polynomial $\varphi_\pi(1,1,...,1)$
of $\tau$ has a fixed parity (see eq.\homopsi\ for the example $2n=6$). 
It appears that this parity is reversed under the action of $e_i$
(when it is non-trivial, i.e. when no chord connects $i$ and $i+1$ in $\pi$),
namely parity$(\varphi_{e_i\pi})=-$parity$(\varphi_\pi)$: this is a direct consequence of
the action \actontwomon, which adds up or removes a $b$, therefore reverses the parity
of the homogeneous quantities.
With the normalization $\varphi_{\pi_0}=1$, this fixes all parities
of the $\varphi_\pi$, say $\epsilon_\pi$. So when $\tau=-1$, we get a new alternating
sum rule
\eqn\alterfpl{ \sum_\pi \epsilon_\pi \, F_n(\pi) = \left\{ 
\matrix{ 0 & {\rm if}\ n\ {\rm is}\ {\rm even}\cr
(-1)^{n-1\over 2} A_V(n)^2 & {\rm if}\ n\ {\rm is}\ {\rm odd}\cr } \right. }
while $\sum_\pi F_n(\pi)=A_n$.

\listrefs

\end